\documentclass[aps,pra,twocolumn,groupedaddress,showpacs]{revtex4}

\usepackage{graphicx}
\usepackage{amsmath}

\begin{document}


\title{Giant hole and circular superflow in a fast rotating Bose-Einstein condensate}


\author{Kenichi Kasamatsu$^1$}
\author{Makoto Tsubota$^1$}
\author{Masahito Ueda$^2$}
\affiliation{$^1$Department of Physics,
Osaka City University, Sumiyoshi-Ku, Osaka 558-8585, Japan \\
$^2$Department of Physics, Tokyo Institute of Technology,  
Meguro-ku, Tokyo 152-8551, Japan}

\date{\today}

\begin{abstract}
A fast rotating Bose-Einstein condensate confined in a 
quadratic-plus-quartic potential is found to dynamically generate 
a ``giant vortex" that absorbs all phase singularities into a central low 
density hole, thereby sustaining a quasi-one-dimensional circular 
superflow at a supersonic speed.
\end{abstract}


\maketitle

Gaseous Bose-Einstein condensates (BECs) are an extremely versatile testing ground 
for superfluidity particularly when an externally driven rotation exists. 
Through a number of experimental \cite{Madison,Abo,Haljan,Hodby} and theoretical 
studies \cite{Fetter1} on rotating BECs, the argument now extends to 
the fast rotating regime \cite{Ho,Fetter2,Feder,Fisher,Lundh}. For a 
harmonic trapping potential $(1/2)m\omega_{\perp}^{2}r^{2}$, the centrifugal 
force prevents a BEC from rotating at frequency $\Omega$ beyond the radial 
trapping frequency $\omega_{\perp}$. 
However, a recent experimental development 
\cite{Kuga} should enable one to create a confinement potential tighter 
than harmonic, thus opening a possible method to explore the nature of 
fast rotating BECs. 

In this article, we extend our previous studies \cite{Tsubota} to 
the dynamics of vortex lattice formation of a 
BEC in a quadratic-plus-quartic potential with $\Omega$ greater 
than $\omega_{\perp}$. Here, the stable vortex configuration, 
realized at a value of $\Omega$ much greater than $\omega_{\perp}$, 
is not triangular but rather a circular array around a central low 
density hole [see Fig. 1(e) below]. This hole is created where singly 
quantized vortices are close together but do not completely overlap. 
Surprisingly, as $\Omega$ is 
further increased, all the vortices are absorbed into the central low 
density hole, around which persistent current circulates.
This superflow is supersonic, which offers an avenue of research  
in superfluidity, because it has been difficult to obtain a  
circular superflow in superfluid helium \cite{Donnely}. 
For an atomic-gas BEC, a circular superflow can occur 
in an ideal situation and the superfluid speed is easily 
controlled by changing $\Omega$. This opens up possibilities 
of studying phase slippage \cite{Anderson} and the stability of 
a controlled superflow. 
\begin{figure}[bp]
\includegraphics[height=0.38\textheight]{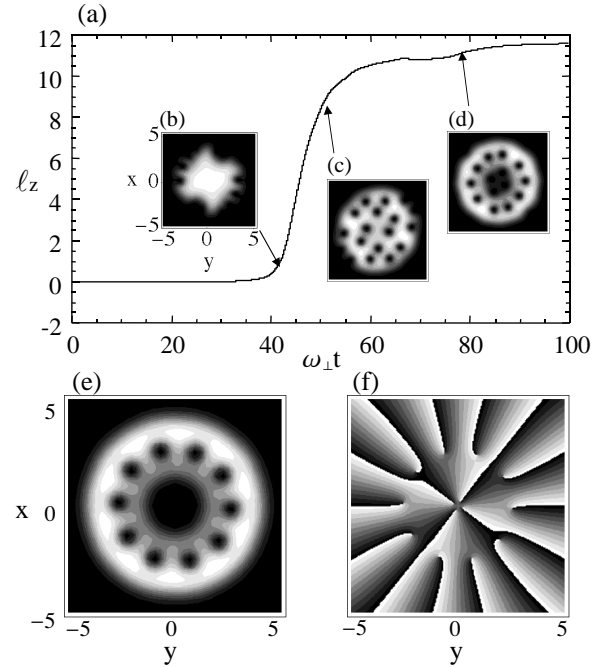}%
\caption{Time development of the angular momentum per atom $\ell_{z}$
(a) and the condensate density $|\Psi|^{2}$ for $\Omega=2.5$ (b)-(e). 
Times $\omega_{\perp} t =$ 42 (b), 50 (c), and 78 (d). The lower 
panel shows the density (e) and phase (f) profiles of the 
final stationary state of this dynamic system. The value of the 
phase varies continuously from $0$ (black) to $2 \pi$ (white). 
The discontinuity lines between black and white are 
the branch cuts in the complex plane. The termination of 
each branch cut is a vortex, or equivalently a phase defects.}
\label{dyna}
\end{figure}

We consider a two-dimensional system subject to rotation 
${\bf \Omega}=\Omega {\bf \hat{z}}$ by assuming translation symmetry 
along the $z$ axis. In a frame rotating with frequency $\Omega$ around the 
$z$ axis, the dynamics of a condensate ``wave function" $\Psi$ 
is described by the Gross-Pitaevskii equation
\begin{equation}
i \frac{\partial \Psi}{\partial t} = \biggl( - \nabla^{2} 
+ V_{\rm trap} - \mu + C |\Psi|^{2}- \Omega \hat{L}_{z} 
\biggr) \Psi, \label{gpe1}
\end{equation} 
where $\mu$ is the chemical potential and the wave function 
is normalized to unity: $\int dx dy |\Psi|^{2} = 1$. 
The units of energy, length, and time are given by the corresponding 
scales of the harmonic potential, that is, $\hbar \omega_{\perp}$, 
$b_{\perp}=\sqrt{\hbar/2m\omega_{\perp}}$, and $\omega_{\perp}^{-1}$, 
respectively, where $m$ is the atomic mass and $\omega_{\perp}$ 
the frequency of the harmonic potential. The strength of the mean-field 
interaction between atoms is characterized by $C$ which is proportional to 
the $s$ wave scattering length $a_{s}$ as $C=8 \pi N_{\rm 2D} 
a_{s}$, where $N_{\rm 2D}$ is the particle number per unit 
length along the $z$ axis. The trapping potential 
is $V_{\rm trap}(r) = r^{2}/4  + 
k r^{4}/16$, where $r^{2} \equiv x^{2}+y^{2}$ and $k$ is 
the strength of the quartic term. The quartic part of $V_{\rm trap}$ 
allows us to increase $\Omega$ above $\omega_{\perp}$. For a rotating 
condensate an effective potential in the radial direction is  
\begin{equation}
V_{\rm trap}^{\rm eff}(r)= \frac{1}{4} \biggl[ 
(1-\Omega^{2}) r^{2} + \frac{1}{4} k r^{4} \biggr]. 
\label{Mexic}
\end{equation}
This potential has a ``Mexican hat" structure for $\Omega > 1$, which 
changes the dynamics of vortex lattice formation as shown below.

Now, we investigate the dynamics of vortex lattice formation \cite{numer}. 
The stationary solution of Eq. (\ref{gpe1}) for a nonrotating trap 
is used as the initial state. The rotating drive is suddenly turned 
on by introducing a small anisotropy of the harmonic trap as 
$( \epsilon_{x} x^{2} + \epsilon_{y} y^{2} )/4$ \cite{Madison,Tsubota}. 
Because the transition from a nonvortex state to a vortex state requires 
energy dissipation, a phenomenological damping constant $\gamma$ is 
put in Eq. (\ref{gpe1}) by replacing $i\partial/\partial t$ with 
$(i-\gamma)\partial/\partial t$ \cite{Tsubota,Choi}. 
Finally, the small trap anisotropy 
$\epsilon_{x,y}$ is turned off adiabatically to obtain the 
axisymmetric stationary state. 

Numerical simulations were done for several values 
of $\Omega$ at fixed $C=250$ and $k=1$. Dynamics that are unique to 
the quadratic-plus-quartic potential occur when $\Omega>1$. 
Figure \ref{dyna} shows the time 
development of the angular momentum per atom $\ell_{z} = \int 
dx dy \Psi^{\ast} \hat{L}_{z} \Psi$, and the density and phase profiles 
for $\Omega=2.5$. The ripples are excited on the surface [Fig. \ref{dyna}(b)], 
and some ripples penetrate into the condensate and 
develop into vortex cores [Fig. \ref{dyna}(c)]. 
Some penetrating vortices move toward the center, merging together 
to make a hole at the center [Fig. \ref{dyna}(d)]. 
During the period between that shown in Fig. \ref{dyna}(c) 
and that in Fig. \ref{dyna}(d), additional vortices entered the 
condensate successively, which gradually increased the angular momentum. 
The density and phase profiles of the final steady state are shown 
in Figs. \ref{dyna}(e) and \ref{dyna}(f), respectively. 
A hole appears at the center, around which some vortices form 
a circular array. Furthermore, near the center of the condensate, 
some phase defects come very close to each other without overlapping 
[Fig. \ref{dyna}(f)].

\begin{figure}[ht]
\includegraphics[height=0.2\textheight]{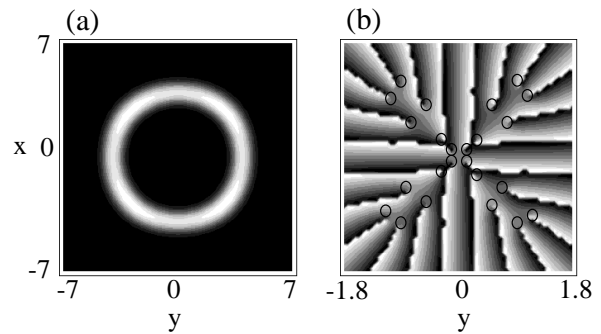}%
\caption{The condensate density (a) and phase field (b) of 
the vortex state with circulation $24 h/m$ for $\Omega=3.2$.
The phase field region in (b) is the central low density region of (a), 
where 24 phase defects are marked by open circles.}
\label{grond}
\end{figure}

As $\Omega$ is further increased, all vortices generated 
from the condensate surface are absorbed into the central low 
density hole as shown in Fig. \ref{grond}(a). Note that this central hole 
is composed of 24 singly quantized vortices packed together [Fig. \ref{grond}(b)]. 
This packing is possible for high rotation frequencies ($\Omega>1$) because 
the centrifugal force decreases the condensate density near the center; 
thus packing vortices costs little energy.
The minimum of Eq. (\ref{Mexic}) determines the radius of the ring 
condensate in Fig. \ref{grond} as $R=\sqrt{2(\Omega^{2}-1)/k}$. 
Here we call a set of vortices such as that in Fig. \ref{grond} a 
``giant vortex" to indicate that a number of phase defects 
are contained in a single hole [see Fig. \ref{grond}(b)].
Such a density depression with a concentration of vorticity occurs 
in rotating superfluid helium \cite{Marston}. 

Figures \ref{dyna}(f) and \ref{grond}(b) show that the phase
singularities do not completely overlap. 
The condensate inside the hole has a very low density and hence 
a long coherence length, so that the vortex cores have a nearly uniform 
density. Thus, only the locations of 
singularities matter and one can assume that the 
interaction energy between the singularity points is 
proportional to $-\rho \kappa^{2} \ln d$, where $\rho$ 
is the condensate density, $\kappa$ the quantum circulation, 
and $d$ the distance between defects. The logarithmic 
divergence at $d \rightarrow 0 $ prevents the phase 
singularities from merging together. 

The appearance of such a density hole is relevant to recent studies on fast 
rotating BECs that used the Thomas-Fermi and Wigner-Seitz approximation 
\cite{Fetter2,Fisher}. Fischer and Baym evaluated the vortex core size self-consistently 
by minimizing the Gross-Pitaevskii energy functional and using a variational wave function 
\cite{Fisher}; they showed that vortex cores never overlap, in agreement with our finding. 
They also showed that the giant vortex state is energetically favorable at large $\Omega$. 
Although the Thomas-Fermi approximation accurately describes stationary solutions 
such as that in Fig. \ref{dyna}(e), it fails when the width of the 
ring condensate becomes thin such as that in Fig. \ref{grond}. 
Even if $\Omega$ is not large enough, a pinning potential might be used to 
stabilize the giant vortex \cite{Simula}. 

For sufficiently large $\Omega$, the ring structure of 
Fig. \ref{grond}(a) offers some insights into circular 
superflow, which has been studied for a torus \cite{Donnely}. 
The superfluid velocity is given by the gradient of the 
phase of the wave function as ${\bf v}_{s}=(\hbar/m) \nabla 
\theta$; the circulation along a closed loop is quantized 
as $\oint {\bf v}_{s} \cdot d{\bf l} = \kappa n $ with $\kappa \equiv h/m$ being the 
quantum circulation. The quantum number $n$ is equal to the 
number of branch cuts of the phase field along the loop or, 
equivalently, the number of phase defects in the low density hole. 
If there is sufficient dissipation, some vortices in the low density 
hole should spiral out of the condensate when $\Omega$ is abruptly 
decreased from that of Fig. \ref{grond}(a). 
The motion of these vortices 
across the circular superflow causes $v_{s}$ to decay. 
This is known as phase slippage \cite{Anderson}, which is a characteristic 
feature of superfluidity. 

\begin{figure}[ht]
\includegraphics[height=0.31\textheight]{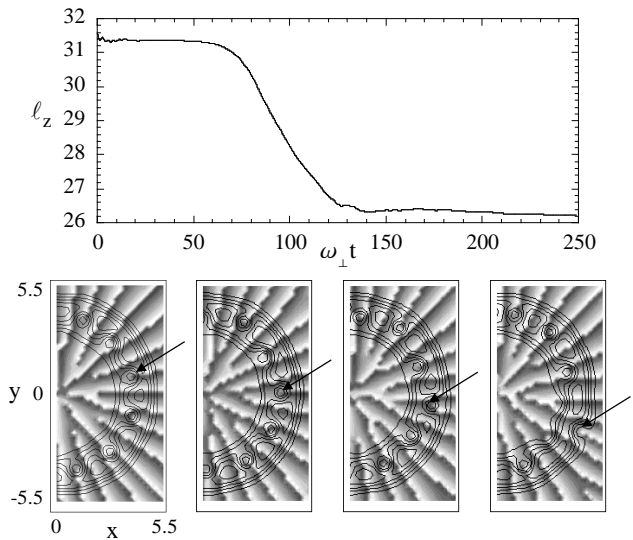}%
\caption{The top figure shows the time evolution of the angular 
momentum per atom $\ell_{z}$ for a change of $\Omega$ from 3.5 to 3.2. 
The bottom figures show the phase profiles and line contours of the density 
for half of the $x-y$ plane at 
$\omega_{\perp}t=102,107,112$, and $116$ from left to right. 
Arrows show how one vortex escapes via phase slippage.}
\label{angl}
\end{figure}

Measurement of the angular momentum of the giant vortex allows the 
phase slippage to be observed well under control. For example, phase 
slippage was observed in helium superflow 
through microapertures \cite{Varoquaux}, but it was 
difficult to control single phase slips and giant slips. 
Figure \ref{angl} shows the time development of the angular momentum and 
phase field resulting from a decrease in $\Omega$ from 
$\Omega=3.5$ ($n=32$) to $\Omega=3.2$. Here we assumed a 
small anisotropy in the trapping frequencies to break 
the rotational symmetry. Some phase defects are released 
from the low density hole, jostling at the inner border of 
the ring condensate. We see that four vortices cross 
the superflow by phase slippage and then escape 
outside (only one escape is shown in Fig. \ref{angl}), 
thus decreasing the angular momentum.

When the ring radius $R$ is sufficiently large, the circular 
flow becomes quasi-one-dimensional for the 
following reason. The radial wave function of the stationary 
giant vortex state is obtained by substituting 
$\Psi \cong f(r)e^{in\theta}$ into Eq. (\ref{gpe1}):
\begin{equation}
\biggl[ -\frac{\partial^{2}}{\partial r^{2}}-\frac{1}{r}
\frac{\partial}{\partial r}+\frac{n^{2}}{r^{2}}
+V_{\rm trap}+C|f|^{2}-n\Omega \biggr] f = \mu f.
\end{equation}
The excitation of the radial component is easily obtained 
by writing $f+\delta f$; the effective 
potential for $\delta f$ is then given by 
\begin{equation}
V_{\rm eff}(r)=V_{\rm trap}+\frac{n^{2}}{r^{2}}
+2 C |f(r)|^{2} - \mu -n\Omega.
\end{equation}
Because $n=\Omega R^{2}/2$ in the limit of a rigid 
body rotation \cite{Fetter2,Feder}, the radial zero-point energy 
around the minimum of the potential $V_{\rm trap}+n^{2}/r^{2}$ 
becomes larger than the mean-field interaction strength 
$C|f(r)|^{2}$ for large $R$. Such a high excitation gap freezes 
the motion of the radial wave function while keeping the motion 
free along the $\theta$ direction. To investigate 
the stability of this one-dimensional superflow, 
we substitute $\Psi(r, \theta, t)=f(r) 
\psi(\theta,t)$ into Eq. (\ref{gpe1}) to obtain
\begin{equation}
 i \frac{\partial \psi}{\partial t} = \biggl( - \frac{1}{R^{2}} 
 \frac{\partial^{2}}{\partial \theta^{2}} + E_{r} 
 + C P |\psi|^{2}- i \Omega \frac{\partial}
 {\partial \theta} \biggr) \psi, \label{gpe1dim}
\end{equation}
where
\begin{eqnarray}
E_{r}= \frac{1}{A} \int r dr f^{\ast} \biggl[ 
-\frac{\partial^{2}}{\partial r^{2}}-\frac{1}{r}
\frac{\partial}{\partial r}+V_{\rm trap} - \mu \biggr]f, 
\nonumber \\
R^{-2} \simeq \frac{1}{A} \int r dr |f|^{2} r^{-2}, 
\hspace{5mm} P= \frac{1}{A} \int r dr |f|^{4} \nonumber
\end{eqnarray}
with $A=\int r dr |f|^{2}$. The wave function is linearized 
as $\psi(\theta, t)= e^{-i \mu t} \{ e^{i n \theta} + \eta 
[ e^{-i \omega t} u(\theta) + e^{i \omega^{\ast} t} 
v^{\ast}(\theta) ] \}$. The resulting eigenvalue problem 
with the ansatz $u(\theta)=u_{q}e^{i(n+q)\theta}$ and 
$v(\theta)=v_{q}e^{i(-n+q)\theta}$ yields the dispersion relation
\begin{equation}
\omega_{q, \pm} = \frac{q}{R} \biggl\{ \biggl(\frac{2n}{R} 
- R \Omega \biggr) \pm \sqrt{ \biggl( \frac{q}{R} \biggr)^{2}+2CP} 
\biggr\},
\label{disper}
\end{equation}
where $q/R$ is the wave number of the excitation. 
The second term on the right-hand side of Eq. (\ref{disper}) gives the 
sound velocity $c=\sqrt{2CP}$. The superflow is stable when $v_{s}<c$ 
according to the Landau criterion. The values of the 
parameters and the form of $f(r)$ of Fig. \ref{grond}(a) 
yield $c=3.75$ and the circular flow velocity $v_{s}=2n/R 
\simeq 12$ (in units of $b_{\perp} \omega_{\perp}$). 
Because $v_{s}>c$, the circular flow is supersonic and energetically unstable. 

\begin{figure}[ht]
\includegraphics[height=0.20\textheight]{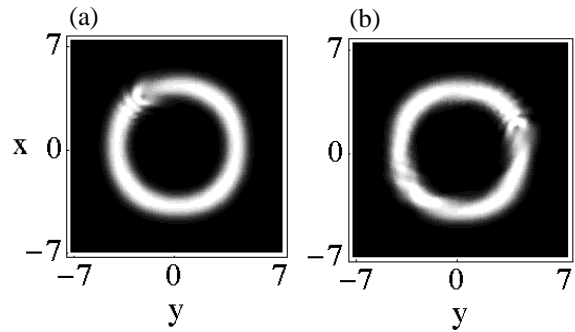}%
\caption{Snapshots of the numerical simulation of a 
giant vortex state (Fig. \ref{grond}) that is perturbed by an 
impurity potential. Parameter values for 
$V_{\rm imp}$ are $V_{0}=100$, $\sigma=0.1$, $x_{0}=3.94$ 
and $y_{0}=0$. The impurity potential is at rest 
in the laboratory frame, and thus rotates in the rotating frame.}
\label{landau}
\end{figure}
Impurities and walls can cause dissipation for a supersonic superflow. 
For example, Ref. \cite{Jackson} used the Gross-Pitaevskii equation 
for bulk condensates and ordinary trapped condensates to show that 
the dissipation of the superflow is caused by vortex-pair creation 
at an impurity object. This was shown experimentally \cite{Raman}. 
However, this does not apply to the present case of 
quasi-one-dimensional superflow because of the tight radial 
confinement \cite{omake}. 
To show the influence of impurities, 
we introduce an impurity potential of Gaussian form as 
$V_{\rm imp}=V_{0} \exp[-\{ (x-x_{0})^{2}+(y-y_{0})^{2} 
\}/\sigma^{2}]$ (at rest in the laboratory frame), where 
$\{x_{0}, y_{0}\}$ is the impurity position. 
Figure \ref{landau} shows an example of the destabilization 
process of the giant vortex state by $V_{\rm imp}$, 
for which $V_{0}$ is chosen to be larger than $V_{\rm trap}(R)$ 
and $\sigma$ smaller than the width of the ring condensate. 
Because $v_{s}$ exceeds the Landau critical velocity $c$, 
the impurity excites the longitudinal density waves along 
the ring condensate, thus producing a drag force. 
Due to the presence of dissipation and the rotating drive, 
the system reaches a steady state as shown in Fig. \ref{landau}(b). 
Therefore, the quadratic-plus-quartic system offers a scheme for studying the 
stability of a quasi-one-dimensional superflow. 

In conclusion, a fast rotating BEC in a quadratic-plus-quartic potential 
can generate a giant vortex and circular superflow around 
it by absorbing singly quantized vortices into a central low density hole. 
The circular superflow is quasi-one-dimensional and goes at a supersonic speed. 
We thus hope that this system will encourage experiments to 
study low dimensional superfluidity in atomic-gas BECs.

M.T. acknowledges support by a Grant-in-Aid for Scientific Research
(Grant No. 12640357) by the Japan Society for the Promotion of Science.
M.U. acknowledges support 
by a Grant-in-Aid for Scientific Research
(Grant No. 11216204) by the Ministry of Education, Science, Sports,
and Culture of Japan, and by the Toray Science Foundation.



\begin{thebibliography}{99}
\bibitem{numer}
The numerical simulations were done with an alternating-direction 
implicit method in the $256 \times 256$ discretized space with the spatial region 
$[-7,7] \times [-7,7]$.
\end{thebibliography}
\end{document}